\documentclass[%
 reprint,
 amsmath,amssymb,
 aps,
]{revtex4-2}

\usepackage{graphicx}% Include figure files
\usepackage{dcolumn}% Align table columns on decimal point
\usepackage{bm}% bold math
\usepackage[makeroom]{cancel}
\usepackage{url}

\usepackage{breakurl}
\usepackage[breaklinks]{hyperref}
\usepackage{graphicx}
\usepackage{xcolor} % add hypertext capabilities
\usepackage{esint} %for slashed integral symbol \fint

\begin{document}

\preprint{APS/123-QED}
\title{Scattering and bound observables for spinning particles in Kerr spacetime \\ with generic spin orientations} 

\author{Riccardo Gonzo}\email{rgonzo@ed.ac.uk} 
\affiliation{Higgs Centre for Theoretical Physics, School of Physics and Astronomy, 
University of Edinburgh, EH9 3FD, UK}%
\author{Canxin Shi}\email{shicanxin@itp.ac.cn}
\affiliation{CAS Key Laboratory of Theoretical Physics, Institute of Theoretical Physics,
Chinese Academy of Sciences, Beijing 100190, China}%

\newcommand{\dd}{\mathrm{d}}
\newcommand{\dx}{\mathrm{d} x}
\newcommand{\calO}{\mathcal{O}}
\newcommand{\calE}{\mathcal{E}}

\begin{abstract}
We derive the radial action of a spinning probe particle in Kerr spacetime from the worldline formalism in the first-order form, focusing on linear in spin effects. We then develop a novel covariant Dirac bracket formalism to compute the impulse and the spin kick directly from the radial action, generalizing some conjectural results in the literature and providing ready-to-use expressions for amplitude calculations with generic spin orientations. This allows, for the first time, to find new covariant expressions for scattering observables in the probe limit up to $\mathcal{O}(G^6 s_1 s_2^4)$. Finally, we use the action-angle representation to compute the fundamental frequencies for generic bound orbits, including the intrinsic spin precession, the periastron advance and the precession of the orbital plane.
\end{abstract}

\maketitle

\section{Introduction}\label{sec:intro}

The discovery of gravitational waves by the LIGO-Virgo-KAGRA collaboration opened up a new avenue for the investigation of the properties of compact astrophysical objects --black holes and neutron stars-- in our universe. Given current and future experiments, the need to develop accurate theoretical templates for the gravitational wave signal is more pressing than ever. In particular, the analytic description of the evolution of compact binaries becomes essential for extreme mass ratio inspirals (EMRIs), which will be one of the main targets for the upcoming space-based detector LISA \cite{Barausse:2020rsu,LISAConsortiumWaveformWorkingGroup:2023arg}.

The classical two-body problem in the mass ratio expansion is naturally described with the gravitational self-force (GSF) approach \cite{Poisson:2011nh,Barack:2018yvs,Pound:2021qin}, which at zeroth order (0SF) becomes equivalent to computing the geodesic solution for the probe particle in the gravitational background generated by the heavy source. The analytic solution of this problem is well-known for spinless particles in Kerr spacetime \cite{Chandrasekhar:1985kt}; much less is known instead for spinning ones due to difficulties in solving the classical Mathisson-Papapetrou-Dixon equations \cite{Mathisson:1937zz,Papapetrou:1951pa,Dixon:1964cjb,Dixon:1970zz,Dixon:1974xoz,Tanaka:1996ht,Hartl:2003da,Hackmann:2014tga,Ruangsri:2015cvg,Bini:2017slb,Bini:2017pee,Witzany:2018ahb,Drummond:2022xej,Drummond:2022efc,Witzany:2023bsa,Compere:2023alp}. This is potentially relevant for self-force inspirals codes \cite{Drasco:2003ky,Hinderer:2008dm,Barausse:2009aa,LeTiec:2013uey,Hinderer:2013uwa,VanDeMeent:2018cgn,vandeMeent:2019cam,Mathews:2021rod,Witzany:2023bmq,Blanco:2023jxf}, where an analytic expression of the spin curvature correction to the fundamental frequencies of the motion would be desirable for generic bound orbits. An alternative perspective on this problem comes from the recently developed effective field theory approach, either amplitude-based or worldline-based \cite{Goldberger:2004jt,Porto:2005ac,Levi:2015msa,Kosower:2018adc,Liu:2021zxr,Jakobsen:2021zvh,Bellazzini:2022wzv,DiVecchia:2022piu,DiVecchia:2023frv,Alessio:2023kgf,Scheopner:2023rzp,Ben-Shahar:2023djm}, where new tools have been produced in the Post-Minkowskian (PM) expansion for the description of spinning particles to derive the conservative Hamiltonian \cite{Vines:2016unv,Vines:2017hyw,Chung:2019duq,Chung:2020rrz,Chen:2021kxt,FebresCordero:2022jts,Aoude:2022thd,Jakobsen:2022fcj,Jakobsen:2023ndj,Aoude:2023vdk,Bern:2022kto,Kosmopoulos:2021zoq} and related scattering observables \cite{Siemonsen:2019dsu,Maybee:2019jus,Guevara:2018wpp,Guevara:2019fsj,Bautista:2021wfy,Jakobsen:2022zsx,Damgaard:2022jem,Menezes:2022tcs,Bianchi:2023lrg,Bautista:2023szu}. In particular, a novel relation between the two-body classical amplitude and the scattering radial action has been found in the conservative regime \cite{Bern:2021dqo,Kol:2021jjc,Adamo:2022ooq,Damgaard:2021ipf}, from which the impulse and spin kick were derived through some conjectural relations in the PM expansion \cite{Bern:2020buy,Cristofoli:2021jas,Kosmopoulos:2021zoq,Gatica:2023iws,Luna:2023uwd}. This raises the questions of whether we can establish a rigorous relation connecting observables and the action derived from amplitudes or the worldline model, perhaps also extending the boundary to bound map \cite{Kalin:2019rwq,Kalin:2019inp,Adamo:2024oxy} found at the non-perturbative level for generic Kerr orbits \cite{Gonzo:2023goe}.

In this paper, we develop new methods to compute scattering and bound observables for geodesics of spinning particles in Kerr spacetime at linear in spin order. First, using the worldline formalism, we show how the radial action in the first-order form effectively provides a Hamiltonian for the evolution of our observables which we can analytically integrate for our model. We will then derive new Dirac brackets to compute the impulse $\Delta v_1^{\mu}$ and the spin kick $\Delta s_1^{\mu}$ for the constrained Hamiltonian system, which will demonstrate the relation between the radial action and the corresponding observables in a covariant fashion. We will use such new relations to compute such scattering observables for the spinning probe up to $\mathcal{O}(G^6 s_1 s_2^4)$. Finally, we will derive the fundamental frequencies using the action-angle representation, including the intrinsic spin precession $K^{\phi_S r}$, the periastron advance $K^{\phi r}$ and the orbital plane precession ${K^{\phi r}-K^{\theta r}}$.

\paragraph*{Conventions} We set $c=G_N=1$ and we use the mostly plus $(-+++)$ signature convention for the metric. %{We also assume $L>0$ throughout the paper.}

\section{The radial action from the first order worldline formalism}
\label{sec:radial_action}

As a warm up, we consider a spinless probe particle moving in a generic spacetime which can be described with the first-order worldline action
\begin{align}
    \label{eq_worldlineAction}
    S[x^\mu(\tau), p_\mu(\tau)] = \int \dd \tau\,  p_\mu \dot{x}^\mu - \frac{e}{2} (g^{\mu\nu} p_\mu p_\nu + m^2) \,,
\end{align}
where $(x^\mu,p_\mu)$ are canonically conjugate coordinate and momentum variables, $e$ is the einbein imposing the on-shell constraint and $g_{\mu \nu}$ is the background metric. We take such metric to be the Kerr spacetime in Boyer-Lindquist coordinates $(t, r, \theta, \phi)$,
\begin{align}
\mathrm{d} s^2= & -\frac{\Delta}{\Sigma}\left(\mathrm{d} t-a \sin ^2(\theta)  \, \mathrm{d} \phi\right)^2+\frac{\Sigma}{\Delta}  \, \mathrm{d} r^2+\Sigma \, \mathrm{d} \theta^2 \nonumber  \\
& +\frac{\sin ^2(\theta)}{\Sigma}\left[\left(r^2+a^2\right) \mathrm{d} \phi-a \mathrm{~d} t\right]^2 \,, \nonumber \\
\Delta(r)= & r^2-2 M r+a^2\,, \quad \Sigma(r, \theta)=r^2+a^2 \cos ^2(\theta) \,,
\end{align}
where $M$ is the mass and $a$ is the radius of the ring singularity.  We now want to parametrize the motion with the radial coordinate $r$, obtaining
\begin{align}
    \label{eq:radial_action_red}
    & S^{\text{eff}}[x^I(r), p_I(r)] \nonumber \\
    &\quad = \int_{\mathcal{C}_r} \dd r \Big[\sum_{I = t, \theta, \phi} p_{I} \frac{\dd x^I}{\dd r}  + p_r(r, x^I(r), p_I(r))\Big] \,,
\end{align}
where the integration follows the physical radial trajectory $\mathcal{C}_r$,
i.e. in general it is not monotonically increasing. To obtain an explicit expression for the radial action, we can use the symmetries of the system. The Kerr metric admits two conserved Killing vectors  $\xi_t^\mu$, $\xi_{\phi}^\mu$ and one Killing tensor $K_{\mu \nu} = Y_{\mu\rho} Y_{\nu}{}^{\rho}$ which is function of the antisymmetric Killing-Yano tensor $Y_{\mu \sigma}=-Y_{\sigma \mu}$
\begin{align}
    \label{eq_KYinln}
    Y_{\mu\nu} \!= a \cos \theta (\ell_\mu n_\nu {-} \ell_\nu n_\mu) + {r\, \epsilon_{\mu\nu\rho\sigma}\ell^\rho n^\sigma} \!\sqrt{{-}\det g_{\mu\nu}}
\end{align}
written in terms of the principal null directions $\ell^\mu, n^\mu$,
\begin{align}
\hspace{-2pt}\ell^\mu=\frac{1}{\Delta}(r^2{+}a^2, \Delta, 0, a),\,\, n^\mu=\frac{1}{2 \Sigma}(r^2{+}a^2,-\Delta, 0, a).
\end{align}
and of the flat space Levi-Civita tensor $\epsilon_{\mu\nu\rho\sigma}$ defined with $\epsilon_{0123} = +1$. A non-trivial property of $Y_{\mu \sigma}$ is that this defines a vector $\mathcal{Y}^{\mu} = Y^{\mu \nu} p_{\nu}$ (besides $p^{\mu}$) which is parallel transported along the geodesic trajectory, i.e. $p^{\nu} \nabla_{\nu} \mathcal{Y}^{\mu} = 0$. This translates into the conservation of the energy $E^{\mathrm{G}}$, the azimuthal component of the angular momentum $L^{\mathrm{G}}$ and the Carter constant $K^{\mathrm{G}} = \mathcal{Y}^{\mu} \mathcal{Y}_{\mu}$ \cite{Carter:1968ks}
\begin{gather}
\label{eq:conserved-charges_spinless}
E^{\mathrm{G}} =-p_\mu \xi_t^\mu=-p_t\,, \quad L^{\mathrm{G}}=p_\mu \xi_{\phi}^\mu=p_\phi\,,\quad \\
%Q^{\mathrm{G}} &=K_{\mu \nu} p^\mu p^\nu-(L^{\mathrm{G}}-a E^{\mathrm{G}})^2 \nonumber \\
K^{\mathrm{G}} \!=\!p_\theta^2 + (a p_t {+} p_\phi)^2 +a^2 (m^2{-}p_t^2 ) \cos ^2(\theta)+p_\phi^2 \cot ^2(\theta)\,. \nonumber 
\end{gather}
Therefore $p_r$ in \eqref{eq:radial_action_red} can be expressed purely in terms of $E^{\mathrm{G}}$, $L^{\mathrm{G}}$ and $K^{\mathrm{G}}$, recovering the well-known radial action
\begin{equation}
%I^{\mathrm{G}}_r(r)  = \int_{\mathcal{C}^{\mathrm{G}}_r} \dd r \,\frac{1}{\Delta} \Big\{ \left[E^{\mathrm{G}}\left(r^2+a^2\right)-a L^{\mathrm{G}}\right]^2 \\
%-\Delta\left(K^{\mathrm{G}}+m^2 r^2\right) \Big\}^{\frac{1}{2}} \,.
I^{\mathrm{G}}_r(r)  = \int_{\mathcal{C}^{\mathrm{G}}_r} \!\!\frac{\dd r}{\Delta}\! { \sqrt{\left[E^{\mathrm{G}} (r^2 {+} a^2){-}a L^{\mathrm{G}} \right]^2 
       - \Delta(K^{\mathrm{G}} + m^2 r^2) }}\,.
\end{equation}

We now extend this derivation to a spinning probe in Kerr spacetime at linear in spin order, building on the supersymmetric $\mathcal{N}=2$ worldline model first developed in \cite{Gibbons:1993ap,Bastianelli:2005vk,Bastianelli:2005uy} and recently revisited in \cite{Jakobsen:2021zvh}. The key idea is to model the spin degrees of freedom with complex Grassmann worldline fields $\psi^{a}$, $\bar{\psi}^a$ living in the local tangent space with flat metric $\eta_{a b}$, so that the spin tensor $S^{\mu \nu}$ is
\begin{align}
S^{\mu \nu}=-2 i e_a^\mu e_b^\nu \bar{\psi}^{[a} \psi^{b]} \,,
\end{align}
and the tetrad basis $e_a^\mu$ satisfies
$%\begin{align}
g_{\mu \nu} = e^a_\mu e^b_\nu \eta_{a b}\,.
%\label{eq:tetrad-def} \end{align}
$
For the purpose of this work, we need only the spinning generalization of the first-order form of \eqref{eq_worldlineAction} 
\begin{align}
\label{eq_worldlineAction-spin}
&\hspace{-5pt}S[x^\mu(\tau), p_\mu(\tau), \psi_a(\tau)] \\
&\hspace{-5pt} \, = \int \dd \tau\, \left[p_\mu \dot{x}^\mu +i \bar{\psi}_a \dot{\psi}^a-e H-i \bar{\chi} \mathcal{Q}-i \chi \bar{\mathcal{Q}}-a \mathcal{J}\right] \,,\nonumber 
\end{align}
where the covariant on-shell and SUSY constraints $H$,$\mathcal{Q}$,$\bar{\mathcal{Q}}$ and the U(1) charge $\mathcal{J}$ are \cite{Bastianelli:2005vk,Bastianelli:2005uy,Jakobsen:2021zvh}
\begin{align}
\mathcal{J} & =\bar{\psi}^a \psi_a-q \,, \quad \mathcal{Q} =\psi^a e_a{ }^\mu \pi_\mu \,,  \quad \bar{\mathcal{Q}}  =\bar{\psi}^a e_a{ }^\mu \pi_\mu \,, \nonumber \\
 & \qquad H =\frac{1}{2} \left[g^{\mu \nu} \pi_\mu \pi_\nu + m^2 + \mathcal{O}(S^2) \right] \,,
\label{eq:constraint-spin}
\end{align}
which are all function of the covariant momentum $\pi_{\mu}$ and therefore of the spin connection $\omega_{\mu a b}$
\begin{align}
\hspace{-4pt}\pi_\mu=p_\mu-i \omega_{\mu a b} \bar{\psi}^a \psi^b \,,\, \omega_\mu^{a b}=e_\nu^a\left(\partial_\mu e^{\nu b}+\Gamma_{\mu \lambda}^\nu e^{\lambda b}\right).
\label{eq:covariant-pi}
\end{align}
The choice of the tetrad in \eqref{eq:covariant-pi} is arbitrary, and a wise choice can drastically simplify the description of the dynamics as we will show shortly. We first notice that there is a direct generalization to the conserved charges \eqref{eq:conserved-charges_spinless}. The two Killing vectors $\xi_t^{\mu}$ and $\xi_{\phi}^{\mu}$ corresponds to the conserved quantities \cite{Ruangsri:2015cvg}
\begin{align}
E&=-\pi_\mu \xi_t^\mu+\frac{1}{2} S^{\alpha \beta} \nabla_\beta (\xi_{t})_{\alpha} \,,  \nonumber \\
L&= \pi_\mu \xi_\phi^\mu-\frac{1}{2} S^{\alpha \beta} \nabla_\beta (\xi_{\phi})_{\alpha} \,,
\label{eq:conserved-chargesEL_spin}
\end{align}
where $E^{\mathrm{G}},L^{\mathrm{G}}$ are defined in \eqref{eq:conserved-charges_spinless} while the R\"udiger constant generalizes the Carter one at dipolar order  \cite{rudiger1981conserved,rudiger1983conserved}
\begin{align}
\!\!\! K = K_{\mu \nu} \pi^\mu \pi^\nu {-} 2\, \pi^\mu S^{\rho \sigma} \left(Y^\nu{ }_\sigma \nabla_\nu Y_{\mu \rho} {-} Y^\nu{ }_\mu \nabla_\nu Y_{\sigma \rho}\right)  .
\label{eq:conserved-chargesK_spin}
\end{align}
Moreover, there is one additional conserved constant in the spin sector, the R\"udiger linear invariant
\begin{equation}
    s_{\parallel}= -\frac{1}{2} \frac{\sqrt{{-}\det g_{\mu\nu}} \epsilon_{\mu \nu \lambda \rho} {Y}^{\mu\sigma} \pi_\sigma S^{\nu \lambda} \pi^\rho}{m^3} \,.
\end{equation}

At this point, we would like to find a separable Hamilton-Jacobi action as for the spinless case earlier. The key idea is that the adapted tetrad $e_{a}^{\mu}$ should be chosen so that it is parallelly transported along the geodesic. This suggests \footnote{To be precise, the solution of the parallel transport equation $u^{\nu} \nabla_{\nu} e_{a}^{\mu} = 0$ with $a=0,1,2,3$ requires a rotation of our tetrad basis.} to use the following tetrad, first derived by Marck \cite{Marck:1983,Kamran:1986} 
\begin{align}
\label{eq:Marck-tetrad}  
e^{0}_{\mu} &= \frac{u_{\mu}}{m} \,, \qquad  
e^{3}_{\mu} = \frac{\mathcal{Y'}_{\mu}}{\sqrt{K}} \,,  \qquad  \mathcal{Y'}^{\mu} = Y^{\mu \nu} u_{\nu}\,,\\
e^{1}_{\mu} &= \frac{\left(K g_{\mu \nu}+m^2 K_{\mu \nu}\right) u^\nu}{m \sqrt{\left(K+m^2 r^2\right)\left(K-m^2 a^2 \cos ^2 \theta\right)}} \,, \nonumber \\
e^{2}_{\mu} &= \frac{\left[K_{\mu \nu} + \left(a^2 \cos^2 \theta (1+ {m^2 r^2}/{K}) - r^2 \right) g_{\mu \nu}\right] \mathcal{Y'}^{\nu}}{r a \sqrt{\cos^2 \theta {(K+m^2 r^2)(K-m^2 a^2 \cos ^2 \theta)}/{K}} }\,, \nonumber 
%N_{(1)} &= m \left(K-m^2 r^2\right)\left(K-m^2 a^2 \cos ^2 \theta\right) \,, \nonumber \\
%N_{(2)} &= m \,r^2 a^2 \cos ^2 \theta \frac{(K-m^2 r^2)(K-m^2 a^2 \cos ^2 \theta)}{K} \,.  \nonumber 
\end{align}
where $u^\mu$ corresponds to a fiducial geodesic parametrized by the conserved quantities $E,L,K$ as given in \eqref{eq:conserved-chargesEL_spin} and \eqref{eq:conserved-chargesK_spin}, such that $u_t = -E$, $u_\phi = L$ and $g^{\mu\nu} u_\mu u_\nu = -m^2$\footnote{Technically speaking, this tetrad is not globally defined on the spacetime, but this is not relevant for the dynamics and the observables in this paper, and therefore we do not comment on this any further.}.
Using this tetrad we then obtain from the Hamiltonian constraint at linear in spin order \eqref{eq:constraint-spin} the radial and the polar momentum (in agreement with \cite{Witzany:2019nml})
\begin{align}
\label{eq:radial_polar}
p_r&(r) = \frac{1}{\Delta} \bigg\{ \left(E \left(a^2+r^2\right)-a L\right)^2-\Delta \left(K+m^2 r^2\right) \nonumber \\
&\qquad +\Delta \bigg(\frac{2 m^3 s_{\parallel} \left(E (a^2+r^2)-a L\right)}{ K+m^2 r^2 }\bigg) \bigg\}^{\frac{1}{2}}\,, \nonumber \\
p_{\theta}&(\theta) = \bigg\{K- \Big(\frac{L}{\sin (\theta)}-a E \sin (\theta )\Big)^2 -a^2 m^2 \cos ^2(\theta ) \nonumber \\
&\qquad +\frac{2 a m^3 s_{\parallel} (L-a E \sin ^2(\theta ))}{ K-a^2 m^2 \cos ^2(\theta )}\bigg\}^{\frac{1}{2}}\,.
\end{align}
The fact that the system admits a separable action is a signature of the integrability of the system, as recently pointed out in \cite{Ramond:2022vhj,Ramond:2024ozy,Ramond:2024sfp}. We now apply the radial reduction method explained earlier to the spinning probe action \eqref{eq_worldlineAction-spin}, obtaining
\begin{align}
\label{eq:spinningradial_action_red}
&S^{\text{eff}}[x^I(r), p_I(r), \psi_a(r)] \\
&\,= \int_{\mathcal{C}_r} \!\dd r \Big[\! \sum_{I = t, \theta, \phi}\!\! p_{I} \frac{\dd x^I}{\dd r} +i \bar{\psi}_a \frac{\dd \psi^a}{\dd r} + p_r(r, E, L, K, s_{\parallel})\Big] \,. \nonumber 
\end{align}
As we will see in section \ref{sec:Dirac_brackets}, the consequence of the representation \eqref{eq:spinningradial_action_red} is that we can effectively consider the radial action as the Hamiltonian for the evolution along the physical trajectory $\mathcal{C}_r$, from which all scattering observables can be derived. It will still be useful, though, to consider the polar momentum in \eqref{eq:radial_polar} as the definition of fundamental frequencies in section \ref{sec:bound-obs} will require action-angle variables \cite{Schmidt:2002qk}.

\subsection{Linear in spin integrated radial action}

In our previous work \cite{Gonzo:2023goe}, we derived explicit non-perturbative expressions for the spinless radial action $I_r^{\mathrm{G}}$ in terms of hypergeometric functions both for scattering and bound orbits. Using convenient unit-mass variables 
\begin{gather}
\bar{E}=E/m\,, \quad \mathcal{E}=\bar{E}^2 - 1\,, \quad l=L/m \,, \quad \bar{K}=K/m^2  \nonumber \\
\bar{Q}=Q/m^2 =\bar{K}-(a \bar{E} - l)^2\,,\quad l_Q=\sqrt{l^2 + \bar{Q}} \,, 
\end{gather}
the radial action the countour of integration is of the form
\begin{align}
\hspace{-5pt}\int_{\mathcal{C}^{\mathrm{G} >}} = 2 \int_{r^{\mathrm{G}}_{m}(\mathcal{E}^{\mathrm{G}},l^{\mathrm{G}},a,l^{\mathrm{G}}_Q)}^{+\infty} \,, \, \int_{\mathcal{C}^{\mathrm{G} <}} = 2 \int_{r^{\mathrm{G}}_{-}(\mathcal{E}^{\mathrm{G}},l^{\mathrm{G}},a,l^{\mathrm{G}}_Q)}^{r^{\mathrm{G}}_{+}(\mathcal{E}^{\mathrm{G}},l^{\mathrm{G}},a,l^{\mathrm{G}}_Q)} \,,
\label{eq:countour_radial}
\end{align}
where the scattering turning point $r^{\mathrm{G}}_{m}$ is the maximal real root of $p_r(r)|_{s_{\parallel} = 0}=0$ and the bound roots $r^{\mathrm{G}}_{\pm}$ obey 
\begin{align}
\label{eq:B2B-turningpoints}
r^{\mathrm{G}}_{\mp}(\mathcal{E}, l, a, l_Q) \stackrel{\mathcal{E}<0}{=} r^{\mathrm{G}}_{m} (\mathcal{E},\pm l, \pm a, \pm l_Q) \,.
\end{align}
Returning to the full spinning radial action, we now have for the scattering case in terms of the variable $u=1/r$
\begin{align}
\label{eq:radial_linearspin}
\hspace{-5pt}I_r^{>} &= 2 \int_{r_{m}(\mathcal{E},l,a,l_Q,s_{\parallel})}^{+\infty} \dd r \,\, p_r(r, E, L, K, s_{\parallel}, a) \\
&= 2 m \int_{0}^{u_{m}(\mathcal{E},l,a,l_Q,s_{\parallel})}  \frac{\dd u }{u^2\, \Delta_u} 
    \sqrt{ R^{\mathrm{G}}(u) + s_{\parallel}\, R^{\mathrm{S}}(u)}  \nonumber 
\end{align}
where $\Delta_u = u^2 \Delta (1/u)$ and the radial potentials are
\begin{align}
    \label{eq_radialpotentials1}
     R^{\mathrm{G}}(u) &= a^2 u^2 \left(u^2 \left(l^2-l_Q^2\right)+{\mathcal{E}}\right)-l_Q^2 u^2 \\
    &\quad  +2 M u \left[u^2 \bar{K}+1\right]+{\mathcal{E}}\,, \nonumber \\
     R^{\mathrm{S}}(u) &= {2 \Delta_u u^2}\frac{(a^2 u^2 + 1){\sqrt{\mathcal{E}+1}}-a l u^2}{u^2 \bar{K}+1}\,. \nonumber 
\end{align}
The turning point $u_{m}$ solves $R^{\mathrm{G}}(u) + s_{\parallel}\, R^{\mathrm{S}}(u) = 0$, but at linear in spin level it is sufficient to consider
\begin{align}
u_{m} = u_{m}^{\mathrm{G}} + s_{\parallel} u_{m}^{\mathrm{S}} + \mathcal{O} (s_{\parallel}^2)
\end{align}
so that the radial action \eqref{eq:radial_linearspin} can be expanded as 
\begin{align}
\label{eq:radial_linearspin-final}
I_r^{>} &= I_r^{\mathrm{G} >} + s_{\parallel} I_r^{\mathrm{S} >}\,, \\
I_r^{\mathrm{S} >} &=  2 m \int_0^{ u_{m}^{\mathrm{G}}} \frac{\dd u }{u^2\, \Delta_u} \left[ \frac{R^{\mathrm{S}}(u)}{2 \sqrt{R^{\mathrm{G}}(u)} }\right] \nonumber \\
&\quad + \cancel{{2 m} u_{m}^{\mathrm{S}} \frac{{\sqrt{R^{\mathrm{G}}(u)}}}{u^2 \Delta_u}\bigg|_{u=u_{m}^{\mathrm{G}}}}  \nonumber 
\end{align}
where the last term vanishes because $R^{\mathrm{G}}(u_{m}^{\mathrm{G}}) = 0$, which means that we do not need the linear-in-spin corrections to the turning points! Finally, we can carry out the integration in an explicit form. The first term $I_r^{\mathrm{G} >}$, with an appropriate IR-regulator $\epsilon>0$, was computed in \cite{Gonzo:2023goe} (see eq. 18) using the decomposition of the $R^{\mathrm{G}}(u)$ polynomial
\begin{align}
     R^{\mathrm{G}}(u) = {\mathcal{E}} \left(1-\frac{u}{u_{m}^{\mathrm{G}}}\right) \prod_{j=1}^3 \left(1-\frac{u}{u_{j}^{\mathrm{G}}}\right)
\end{align}
in terms of the roots $\{u_{m}^{\mathrm{G}},u_1^{\mathrm{G}},u_2^{\mathrm{G}},u_3^{\mathrm{G}}\}$. The new linear in spin term can be computed in a similar way, obtaining
\begin{align}
\label{eq:radial_scattering}
    & \frac{I_r^{\mathrm{S} >}}{m} = \frac{ 4\, a (a \bar{E} - l)}{{\bar{K}\sqrt{\mathcal{E}}}} u_{m}^{\mathrm{G}}  
    F_D^{(3)} \left(1, \frac{1}{2},\frac{1}{2},\frac{1}{2}, \frac{3}{2}; \frac{u_{m}^{\mathrm{G}}}{u_1^{\mathrm{G}}}, \frac{u_{m}^{\mathrm{G}}}{u_2^{\mathrm{G}}},\frac{u_{m}^{\mathrm{G}}}{u_3^{\mathrm{G}}} \right) \nonumber \\
    &\qquad +2 \frac{\bar{K} \bar{E} - a (a \bar{E} - l)}{{\bar{K}\sqrt{\mathcal{E}}}} u_{m}^{\mathrm{G}}  \nonumber \\
    &\qquad \quad  \times \Big[ F_D^{(4)} \left(1, \frac{1}{2},\frac{1}{2},\frac{1}{2},1, \frac{3}{2}; \frac{u_{m}^{\mathrm{G}}}{u_1^{\mathrm{G}}}, \frac{u_{m}^{\mathrm{G}}}{u_2^{\mathrm{G}}},\frac{u_{m}^{\mathrm{G}}}{u_3^{\mathrm{G}}}, i \sqrt{\bar{K}} u_{m}^{\mathrm{G}} \right) \nonumber \\
    &\qquad \qquad+ \left(i \sqrt{\bar{K}} \to -i \sqrt{\bar{K}}\right) \Big]\,,
\end{align}
where $F_D^{(n)}$ is Lauricella's hypergeometric function 
\begin{align}
F_D^{(n)}(&\alpha,\{\beta_j\}_{j=1}^n,\gamma;\{x_j\}_{j=1}^n) = \frac{\Gamma(\gamma)}{\Gamma(\alpha) \Gamma(\gamma-\alpha)} \nonumber \\
& \times \int_0^1 \mathrm{d} t \, t^{\alpha - 1} (1-t)^{\gamma - \alpha - 1} \prod_{j=1}^n (1-x_j t)^{-\beta_j} \,.
\end{align}

\section{Covariant Dirac brackets for generic spinning binaries}\label{sec:Dirac_brackets}

In this section, we propose a novel way to extract classical scattering observables from the radial action. \begingroup\color{black}
Before diving into the details, we consider the following relation between the radial action and the conservative 4-point classical amplitude of spinning particles
\endgroup \cite{Bern:2021dqo,Kol:2021jjc,Adamo:2022ooq,Damgaard:2021ipf}
\begin{align}
\mathcal{M}^{\text{cl}}_{4}(q) &=\frac{4 \sqrt{(p_1 \cdot p_2)^2-m_1^2 m_2^2}}{\hbar^{2}} \nonumber \\
& \qquad \times \int \mathrm{d}^{2} b\, e^{-\frac{i}{\hbar} q \cdot b}\left(e^{\frac{i}{\hbar} (I_r^{>} (b) + \pi L)}-1\right) {.}
\label{eq:radial_action_amplitude}
\end{align}
{Here} we use the conventional partial wave decomposition into spherical harmonics (see appendix \ref{app:ampl_action}). The radial action $I_r^{>}$ is a function of the momenta, the spin vectors and the impact parameter $\{p_i^{\mu},s_i^{\mu},b^{\mu}\}$. However, these variables are not independent because a covariant description is highly redundant, requiring proper constraints to remove the unphysical degrees of freedom.

A standard way to treat constrained systems canonically is by Dirac's approach, which boils down to promoting Poisson brackets to Dirac brackets to take care of all second class constraints. 
For scattering events, the radial action can be written completely in terms of the incoming kinematics (see Section~\ref{sec:scattering-obs}), so we only need to impose the constraints on the initial state.
We parametrize the initial straight line particle trajectories $x_1(\tau_1), x_2(\tau_2)$ as
\begin{align}
    x_1^\mu(\tau_1) = v_1^\mu \tau_1 + b_1^\mu, \qquad
    x_2^\mu(\tau_2) = v_2^\mu \tau_2 + b_2^\mu,
\end{align}
with initial spin tensor $S_1^{\mu\nu}$ and $S_2^{\mu\nu}$, respectively. The na\"ive Poisson brackets from a conventional relativistic description of spinning point particles are simply
\begin{gather}
%    \{b_i^\mu, v_i^\nu\}_{_\mathrm{P.B.}} = -\frac{\eta^{\mu\nu}}{m_i},
%    \\
%    \{S_i^{\mu \nu}, S_i^{\alpha \beta}\}_{_\mathrm{P.B.}}= S_i^{\mu \beta} \eta^{\nu \alpha} - S_i^{\mu \alpha}\eta^{\nu \beta} - (\mu \leftrightarrow \nu)
    \{b_i^\mu, v_i^\nu\}_{_\mathrm{P.B.}} = \frac{\eta^{\mu\nu}}{m_i},
    \\
    \{S_i^{\mu \nu}, S_i^{\alpha \beta}\}_{_\mathrm{P.B.}}= S_i^{\mu \alpha}\eta^{\nu \beta} - S_i^{\mu \beta} \eta^{\nu \alpha} - (\mu \leftrightarrow \nu)
    \nonumber
\end{gather}
for $i \in \{1,2\}$, and the other brackets vanish. Regarding the constraints on the spin degrees of freedom, we choose the Tulczyjew spin supplementary condition (SSC)~\footnote{We note that at past infinity $t \to -\infty$, the Tulczyjew SSC coincides with the covariant SSC.}
\begin{align}
    v_{i, \mu} S_i^{\mu\nu} = 0.
\end{align}
Additionally, we need to include proper conjugate gauge constraints, such as $\Lambda_{0\mu}=p_{\mu}/m$, where $\Lambda_{0\mu}$ is the time-like component of the body-fixed frame (see ~\cite{Hanson:1974qy}). The Dirac brackets satisfying these conditions are
\begin{align}
%        & \{b^\mu, v_i^\nu \}^{\prime}= \mathrm{sgn}_i \frac{\eta^{\mu \nu}}{m_i} \nonumber\\
%        & \{b^\mu, b^\nu \}^{\prime}=-S_1^{\mu \nu} / m_1^2 -S_2^{\mu \nu} / m_2^2 \nonumber\\
%        & \{b^\mu, S_i^{\nu \lambda} \}^{\prime}= -\mathrm{sgn}_i \big(S_i^{\mu \nu} v_i^\lambda - S_i^{\mu \lambda} v_i^\nu\big) / m_i \\
%        &\{S_i^{\mu \nu}, S_i^{\alpha \beta}\}^{\prime}=  S_i^{\mu \beta}\big(\eta^{\nu \alpha}+v_i^\nu v_i^\alpha \big) - S_i^{\mu \alpha}\big(\eta^{\nu \beta} + v_i^\nu v_i^\beta \big) \nonumber\\
%        &\qquad\qquad -S_i^{\alpha \nu} \big(\eta^{\mu \beta} + v_i^\mu v_i^\beta \big) + S_i^{\beta \nu} \big(\eta^{\mu \alpha} + v_i^\mu v_i^\alpha \big), \nonumber
    & \{b^\mu, v_i^\nu \}^{\prime}= \mathrm{sgn}_i \frac{\eta^{\mu \nu}}{m_i} \nonumber\\
    & \{b^\mu, b^\nu \}^{\prime}= S_1^{\mu \nu} / m_1^2 + S_2^{\mu \nu} / m_2^2 \nonumber\\
    & \{b^\mu, S_i^{\nu \lambda} \}^{\prime}= -\mathrm{sgn}_i \big(S_i^{\mu \nu} v_i^\lambda - S_i^{\mu \lambda} v_i^\nu\big) / m_i \\
    &\{S_i^{\mu \nu}, S_i^{\alpha \beta}\}^{\prime}= S_i^{\mu \alpha}\big(\eta^{\nu \beta} + v_i^\nu v_i^\beta \big) - S_i^{\mu \beta}\big(\eta^{\nu \alpha}+v_i^\nu v_i^\alpha \big) \nonumber\\
    &\qquad\qquad + S_i^{\alpha \nu} \big(\eta^{\mu \beta} + v_i^\mu v_i^\beta \big) - S_i^{\beta \nu} \big(\eta^{\mu \alpha} + v_i^\mu v_i^\alpha \big), \nonumber
\end{align}
where $b^\mu = b_2^\mu - b_1^\mu$ is the impact parameter and we defined %$\mathrm{sgn}_1 = +1$, $\mathrm{sgn}_2 = -1$.
$\mathrm{sgn}_1 = -1$, $\mathrm{sgn}_2 = +1$.
Furthermore, the freedom of redefining $\tau_1$,$\tau_2$ allows to choose the constraints
\begin{equation}
\begin{split}
    \phi_1=b \cdot v_1 = 0, \quad
    \phi_2=v_1^2 = -1, \\
    \phi_3=b \cdot v_2 = 0, \quad
    \phi_4=v_2^2 = -1,
\end{split}
\end{equation}
Taking advantage of the iterative property, we obtain the Dirac brackets for generic functions $f, g$ by
\begin{align}
\{ f, g \} = \{ f, g \}^{\prime} - \sum_{j,k=1}^{4}\{ f, \phi_j \}^{\prime} (M^{-1})_{j,k} \{ \phi_k, g \}^{\prime},
\end{align}
where $M_{j,k} = \{ \phi_j, \phi_k \}^\prime$.
For efficient calculation, we translate them to be written in terms of the spin vectors defined by $s_i^\mu = - \epsilon^{\mu}{}_{\nu \rho \sigma} v_i^\nu S_i^{\rho\sigma} /(2m_i)$. 
Finally, the Dirac brackets satisfying all the constraints read
\begin{subequations}
    \label{eq_DiracBrackets}
    \begin{align}
        \{b^\mu, v_i^\nu\} &= \mathrm{sgn}_i \frac{(\sigma^2-1)\, b^{\mu } b^{\nu }+\varepsilon^{\mu } \varepsilon^{\nu }}{m_i b^2 \left(\sigma ^2-1\right)},
        \\
        \{b^\mu, s_i^\nu\} &= \mathrm{sgn}_i \frac{v_i^{\nu } \left( (\sigma^2 - 1) b^{\mu } b\cdot s_i+\varepsilon^{\mu } \varepsilon\cdot s_i\right)}{m_i b^2 \left(\sigma ^2-1\right)},  \\
        \{b^\mu, b^\nu\} &= \frac{\left(b^2 (\sigma  v_2^{\mu }-v_1^{\mu })+\varepsilon^{\mu } s_1\cdot v_2\right)b^{\nu } -(\mu \leftrightarrow \nu)}{m_1 b^2 \left(\sigma ^2-1\right)} \nonumber \\
        &\qquad + (1\leftrightarrow 2,\, b \to -b), \\
%        \{s_i^\mu, s_i^\nu\} &= \frac{\mathrm{sgn}_i}{m_i b^2 (\sigma^2-1)}
%        \big(b^{\mu } \varepsilon^{\nu } s_i {\cdot} v_{\bar{i}} + \varepsilon {\cdot} s_i b^{\mu } (\sigma  v_i^{\nu } {-} v_{\bar i}^{\nu} ) \nonumber \\
%        &\qquad - b{\cdot} s_i \, \varepsilon^{\mu } (\sigma  v_i^{\nu} - v_{\bar{i}}^{\nu} ) \big) - (\mu \leftrightarrow \nu),
        \{s_i^\mu, s_i^\nu\} & = \frac{\epsilon^{\mu\nu}{}_{\rho\sigma}  v_i^\rho s_i^\sigma}{m_i} \, ,
    \end{align}
\end{subequations}
where we defined 
\begin{align}
    \label{eq_sigmaeps}
    \sigma = -v_1 \cdot v_2 \, , \qquad
    \varepsilon^\mu = \epsilon{}^\mu{}_{\nu\rho\sigma} b^\nu v_1^\rho v_2^\sigma \, .
\end{align}
Notice that $\varepsilon^\mu$, together with $b^\mu, v_1^\mu, v_2^\mu$, forms a complete vector basis. It is easy to check that the brackets \eqref{eq_DiracBrackets} satisfy the constraints $\phi_j$'s and the spin transversality condition $p_i \cdot s_i =0$. 
Moreover, $\sigma$ is also a constant to the Dirac brackets.
Thanks to these properties, we can impose the constraints before taking derivatives, which drastically simplifies the calculations.

We propose that classical observables, such as the change of velocity $\Delta v_i^\mu$ or spin $\Delta s_i^\mu$, can be extracted by recursively applying Dirac brackets to the {interaction part of the radial action, $\mathcal{I}_r^{>}:=I_r^{>}-I_r^{>}\big|_{G=0}$},
\begin{align}
    \label{eq_nestedDB2}
    \Delta \lambda^\mu =& %\sum_{j=1} \frac{1}{j!}\{ \dots \{\lambda^\mu, \underbrace{I_r^{>}\}, I_r^{>}\} \dots I_r^{>}\}}_{j \text{ times}}
    \sum_{j=1} \frac{1}{j!} \underbrace{ \{{\mathcal{I}_r^{>}}, \{{\mathcal{I}_r^{>}}, ..., \{{\mathcal{I}_r^{>}}}_{j \text{ times}}, \lambda^\mu\} ... \} \}
%    + \dots,
\end{align}
for $\lambda^\mu \in \{v_1^\mu, v_2^\mu, s_1^\mu, s_2^\mu \}$. Given that the spinning particles are treated on equal footings in \eqref{eq_DiracBrackets}, we conjecture and we provide evidence that the above formula \eqref{eq_nestedDB2} can be applied beyond probe limit:
\begin{itemize}
\item For a spinless probe in Kerr, a full proof based on the Hamilton-Jacobi method is in Appendix~\ref{sec:proveRadialformula};
\item  For a linear-in-spin probe, we recover the covariant impulse and spin kick in ~\cite{Jakobsen:2023ndj} at $\mathcal{O}(G^4 s_1^1 s_2^0, G^4 s_1^0 s_2^1)$;
\item For generic spinless binaries beyond the probe limit, we confirm the expressions of the momentum deflection~\cite{Jakobsen:2022psy,Jakobsen:2022fcj,Mogull:2020sak} by applying the brackets to the 3PM conservative \footnote{This does not mean that our method works only for conservative dynamics, and it would be interesting to apply it in presence of radiative effects.} eikonal phase in \cite{Bjerrum-Bohr:2021din,DiVecchia:2023frv};
\item For generic spinning binaries beyond the probe limit, the brackets \eqref{eq_nestedDB2} allow to verify the observables computed in~\cite{Jakobsen:2021zvh} up to $\mathcal{O}(G^2 s_1^1 s_2^1)$.
\end{itemize}
Finally, we stress that the Dirac brackets are independent of the specific model and generally apply to other theories, such as electrodynamics.

\section{Impulse and spin kick for a spinning probe in Kerr}
\label{sec:scattering-obs}

Here we would like to apply \eqref{eq_nestedDB2} to the spinning probe radial action in \eqref{eq:radial_linearspin-final}. For this purpose, we need to write the conserved quantities in terms of the covariant vectors $\{b^\mu, s_i^\mu, v_i^\mu\}$.
Taking advantage of the asymptotic flatness of Kerr spacetime, the Killing vectors at $r\to\infty$ becomes,
\begin{align}
    \xi_{t}^\mu \to v_2^\mu, \quad
    a\, \xi_{\phi}^\mu \to \epsilon^{\mu}{}_{\nu \rho \sigma} v_2^\nu s_2^\rho (x_2^\sigma {-} x_1^\sigma),
\end{align}
where we have taken particle $1$ as the probe and $2$ as the background, e.g., $m_1 = m$ and $m_2 = M$.
To obtain a covariant expression for $Y^{\mu\nu}$ in \eqref{eq_KYinln} at spatial infinity, we compute the leading contribution at $r \to \infty$ 
\begin{align}
    Y^{\mu\nu} = \epsilon^{\mu\nu}{}_{\rho\sigma} v_2^{\rho} (x_2^{\sigma} -x_1^{\sigma}) -s_2^{\nu } v_2^{\mu }+s_2^{\mu } v_2^{\nu}.
\end{align}
Consequently, the relevant conserved quantities in terms of the incoming kinematics, up to $\mathcal{O}(s_1)$, are
\begin{subequations}
\label{eq:kinematics}
\begin{gather}
    \bar{E} = \sigma, \quad
    a = \sqrt{s_2{\cdot} s_2} \\
    a\, l = \varepsilon \cdot s_2 + \sigma\, s_1 \cdot s_2 + s_1{\cdot} v_2\, s_2{\cdot} v_1 \\
    s_{\parallel} = \varepsilon\cdot s_1  - \sigma \, s_1\cdot s_2 - s_1 {\cdot} v_2\, s_2 {\cdot} v_1 \\
    l_Q = \sqrt{(\sigma ^2{-}1)\, b^2 +2\, \sigma\,  \varepsilon{\cdot} s_1- (s_2 {\cdot} v_1 ){}^2-2\, s_1 {\cdot} s_2 },
\end{gather}
\end{subequations}
where $\sigma$, $\varepsilon^\mu$ are defined in \eqref{eq_sigmaeps}.
The radial action written in $\{b^\mu, s_i^\mu, v_i^\mu\}$ can then be obtained by plugging the above relation in \eqref{eq:radial_linearspin-final} and keeping the terms linear in $s_1^\mu$.
We can now extract the impulse and the spin kick by simply applying \eqref{eq_nestedDB2} to such radial action~\footnote{Since we are in the probe limit, we only need to keep the terms of order $1/m_1$.}, getting
\begin{align}
   &\Delta v_1^{\mu} (\{b^\mu, s_i^\mu, v_i^\mu\}) \Big|_{\mathcal{O}(G^6 s_1 s_2^4)} \,,   \nonumber \\ &\Delta s_1^{\mu} (\{b^\mu, s_i^\mu, v_i^\mu\}) \Big|_{\mathcal{O}(G^6 s_1 s_2^4)} \,.
\end{align}
The explicit results are provided in the ancillary file.

\section{Fundamental frequencies for bound orbits}\label{sec:bound-obs}

The integrability of the linear in spin dynamics in Kerr spacetime has a long history \cite{Gibbons:1993ap,Kubiznak:2011ay,Ramond:2022vhj,Ramond:2024ozy,Ramond:2024sfp}, culminated in a recent construction of a separable Hamilton-Hacobi action \cite{Witzany:2019nml}.
%It was recently shown that the linear in spin dynamics in Kerr spacetime form an integrable system \cite{Ramond:2022vhj,Ramond:2024ozy,Ramond:2024sfp}. 
Therefore, given the separability of radial and polar motion in \eqref{eq:radial_polar} and the conserved charges $(E,L,S_{\parallel} = {m s_{\parallel}/\sqrt{\bar K}})$, it is natural to consider the following action variables
\begin{align}
&J_t = -E\,, \qquad J_{\phi} = L \,, \qquad J_{\phi_S} = S_{\parallel} \,, \nonumber \\
& \,\, J_{r} = \frac{1}{2 \pi} \oint p_{r} \dd r \,, \quad J_{\theta} =  \frac{1}{2 \pi} \oint p_{\theta} \dd \theta \,,
\label{eq:action}
\end{align}
{where $\phi_S$ is a $2\pi$-periodic canonical spin angle, with an additive constant in $J_{\phi_S}$ at fixed spin magnitude suppressed. In \eqref{eq:fund_frequencies} and \eqref{eq:frequency_ratios}, the actions are first re-expressed using $s_{\parallel}=S_{\parallel}\sqrt{\bar K}/m$ and $m^2=-2H$; derivatives then hold the other variables in $(H,Q,E,L,S_{\parallel})$ fixed.} The choice of the actions \eqref{eq:action} is justified from many perspectives. The bound radial action in \eqref{eq:action}, for example, is uniquely determined in the aligned-spin case by the integration of the periastron advance (in full agreement with the literature, as we will see). Moreover, it is suggestive to notice that the definition \eqref{eq:action} seems to agree also with the Hamilton-Jacobi action found in \cite{Witzany:2019nml} (see eq.33), following our integration strategy at linear in spin order outlined in \eqref{eq:radial_linearspin-final}. Using the B2B map in \cite{Gonzo:2023goe} we obtain the bound radial action 
\begin{align}
\label{eq:bound_radial}
\hspace{-8pt}J_{r}(\mathcal{E}, l, a, l_Q) = \frac{I_r^{>}(\mathcal{E}, l, a, l_Q) {-} I_r^{>}(\mathcal{E}, -l, -a, -l_Q)}{2 \pi} \,,
\end{align}
which is a consequence of \eqref{eq:B2B-turningpoints} and of the invariance of $p_r$ under $(\mathcal{E}, l, a, l_Q) \to (\mathcal{E},\pm l, \pm a, \pm l_Q)$. For the polar action, using $p_{\theta}$ derived in \eqref{eq:radial_polar} and expanding at linear in spin order as in \eqref{eq:radial_linearspin-final} we obtain 
\begin{align}
J_{\theta} &= \frac{{2}}{\pi} \int_{\pi/2}^{\theta^{\mathrm{G}}_{+}} p_\theta(\theta)\dd \theta \\ %\sqrt{\Theta(\theta)} 
&= \frac{m}{\pi} \int_{0}^{U_+^{\mathrm{G}}} \left[ \frac{ \sqrt{\Theta^{\mathrm{G}}(U)}}{\sqrt{U(1-U)}} + \frac{s_{\parallel} \Theta^{\mathrm{S}}(U)}{2 \sqrt{U(1-U) \Theta^{\mathrm{G}}(U)}} \right] \dd U \,, \nonumber 
\end{align}
where $U=\cos^2(\theta)$, the potentials $\Theta^{\mathrm{G}}(U)$, $\Theta^{\mathrm{S}}(U)$ are
\begin{align}
\Theta^{\mathrm{G}}(U) &= l_Q^2-\frac{l^2}{1-U}+ a^2 {\mathcal{E}} U \,,\nonumber \\
\Theta^{\mathrm{S}}(U) &= \frac{2 a \left(a (U-1) \sqrt{{\mathcal{E}}+1}+l\right)}{a^2 \left({\mathcal{E}}+1-U\right)-2 a l \sqrt{{\mathcal{E}}+1}+l_Q^2}\,,
\end{align}
and $U^{\mathrm{G}}_{\pm}$ are the physical spinless polar turning points \cite{Gonzo:2023goe}
\begin{align}
U^{\mathrm{G}}_{\pm} &= \frac{a^2 \mathcal{E} - l_Q^2 \pm \sqrt{(a^2 \mathcal{E} - l_Q^2)^2 + 4 a^2 \mathcal{E} \bar{Q}}}{2 a^2 \mathcal{E}}\,.
\end{align}
We then obtain the full integrated bound polar action
\begin{align}
\label{eq:bound_polar}
&\frac{J_{\theta}}{m} = \frac{1}{2} \sqrt{\bar{Q} U^{\mathrm{G}}_{+}} F_D^{(2)}\left(\frac{1}{2},1,-\frac{1}{2}, 2; U^{\mathrm{G}}_+, \frac{U^{\mathrm{G}}_+}{U^{\mathrm{G}}_-}\right) \\
&\qquad + \frac{a s_{\parallel} (l {-} a \bar{E}) \sqrt{U_+^{\mathrm{G}}}}{\bar{K} \sqrt{\bar{Q}}} \nonumber \\ 
&\quad \qquad  \times F_D^{(3)}\left(\frac{1}{2},\frac{1}{2},-1,1,1; \frac{U^{\mathrm{G}}_+}{U^{\mathrm{G}}_-}, \frac{a \bar{E} }{a \bar{E}-l} U^{\mathrm{G}}_+,\frac{a^2}{\bar{K}} U^{\mathrm{G}}_+ \right)\,. \nonumber 
\end{align}

Armed with \eqref{eq:bound_polar} and \eqref{eq:bound_radial} and \textit{\emph assuming the independence of the actions} in \eqref{eq:action} for $\vec{s}_1 \nparallel \vec{s}_2 \nparallel \vec{L}^{\mathrm{G}}$, we can then apply the method pioneered by Schmidt \cite{Schmidt:2002qk} which relies on the inversion of the map $J_{\alpha}(H=-m^2/2,Q,E,L,S_{\parallel})$ to define the fundamental frequencies
\begin{align}
\omega_{i} = \frac{\partial H}{\partial J_i} \,, \qquad i=r,\theta,\phi,\phi_S\,.
\end{align}
A direct generalization of the procedure in \cite{Schmidt:2002qk,Gonzo:2023goe} gives
\begin{align}
    \label{eq:fund_frequencies}
     &\hspace{-6pt}\omega_r = \frac{1}{\Omega} \frac{\partial J_{\theta}}{\partial Q}\,,\quad \quad \omega_{\phi} = \frac{1}{\Omega} \bigg(\frac{\partial J_{r}}{\partial Q} \frac{\partial J_{\theta}}{\partial L} - \frac{\partial J_{r}}{\partial L} \frac{\partial J_{\theta}}{\partial Q} \bigg) \,, \nonumber \\
      &\hspace{-6pt}\omega_{\theta} = -\frac{1}{\Omega} \frac{\partial J_{r}}{\partial Q} \,, \quad \omega_{\phi_S} = \frac{1}{\Omega} \bigg( \frac{\partial J_{r}}{\partial Q} \frac{\partial J_{\theta}}{\partial S_{\parallel}} - \frac{\partial J_{r}}{\partial S_{\parallel}} \frac{\partial J_{\theta}}{\partial Q} \bigg) \,, 
\end{align}
with $\Omega = \frac{\partial J_{r}}{\partial H} \frac{\partial J_{\theta}}{\partial Q}-\frac{\partial J_{r}}{\partial Q} \frac{\partial J_{\theta}}{\partial H}$. Interestingly, the frequency $\omega_{\phi_S}$ appears in spin space, which corresponds to the variation of the probe spin over a bound orbit. We now consider the time-independent frequency ratios,
\begin{align}
\label{eq:frequency_ratios}
K^{\theta r} &= \frac{\omega_{\theta}}{\omega_r} =-\frac{\partial J_r/\partial Q}{\partial J_{\theta} /\partial Q} \,, \nonumber \\
K^{\phi r} &= \frac{\omega_{\phi}}{\omega_r} =-\frac{\partial J_r}{\partial L} + \frac{\partial J_r/\partial Q}{\partial J_{\theta}/\partial Q} \frac{\partial J_{\theta}}{\partial L} \,, \nonumber \\
K^{\phi_S r} &= \frac{\omega_{\phi_S}}{\omega_r} =-\frac{\partial J_r}{\partial S_{\parallel}} + \frac{\partial J_r/\partial Q}{\partial J_{\theta}/\partial Q} \frac{\partial J_{\theta}}{\partial S_{\parallel}} \,,
\end{align}
where $K^{\phi r}$ and ${K^{\phi r}-K^{\theta r}}$ {correspond} to the familiar precession of the periastron and of the orbital plane, while $K^{\phi_S r}$ is an intrinsic spin precession related to the drift of a test spinning gyroscope in Kerr \cite{Schiff:1960,Barker:1975,Barker:1979,Bini:2016iym,Akcay:2016dku,Bini:2017slb} (see also \cite{Hinderer:2013uwa}). The calculation of \eqref{eq:frequency_ratios} is relevant in order to extend GSF techniques to include linear in spin effects for the secondary \cite{Drasco:2003ky,Hinderer:2008dm,Barausse:2009aa,LeTiec:2013uey,Dolan:2013roa,Hinderer:2013uwa,VanDeMeent:2018cgn,vandeMeent:2019cam,Mathews:2021rod,Blanco:2023jxf}. Using \eqref{eq:frequency_ratios}, \eqref{eq:bound_polar},\eqref{eq:bound_radial} and \eqref{eq:radial_linearspin-final} the frequency ratios are then evaluated analytically. We stress again here that these are valid for generic spin orientations and the (degenerate) aligned-spin case needs to be discussed separately, as we now show.

In the equatorial limit where $\vec{s}_1 \parallel \vec{s}_2 \parallel \vec{L}^{\mathrm{G}}$, the actions \eqref{eq:action} become dependent from each other since \eqref{eq:kinematics} gives
\begin{align}
l &= l^{\mathrm{G}}+ a_1 \sigma\,,\quad s_{\parallel} = a_1 (l^{\mathrm{G}} - \sigma a_2)\,, \quad a_i = \sqrt{s_i \cdot s_i}\,, \nonumber \\
& \qquad l_Q = \sqrt{(l^{\mathrm{G}})^2 + 2 \,\sigma\, l^{\mathrm{G}} a_1 - 2 \, a_1 a_2 }\,,
\label{eq:equatorial}
\end{align}
which means that in such case only $J_t$,$J_r$ and $J_\phi$ are relevant. We can then compute the aligned-spin periastron advance $K^{\phi r}$ up to $\mathcal{O}(G^6 s_1 s_2^4)$
\begin{align}
 K^{\phi r }(\sigma,l^{\mathrm{G}},a_1,a_2) \big|_{\mathcal{O}(G^6 s_1 s_2^4)}\,,
 \label{eq:periastron}
\end{align}
whose explicit result is provided in the ancillary file. We find agreement with \cite{Liu:2021zxr} up to $\mathcal{O}(G^2 s_1 s_2^2)$ in the probe limit, as well as with the PM expansion of \cite{Hackmann:2014tga}. Since only the radial action contributes in this limit, we notice that this can  also be derived through the standard analytic continuation of the equatorial scattering angle \eqref{eq:bound_radial}.

\section{Conclusions and future directions}\label{sec:conclusion}

\begin{table}[t!]
\begin{tabular}{|c|c|c|}
\hline Type of observable & Position space & Spin space \\
\hline Scattering &  $\Delta v_1^{\mu}$ ($\Delta \phi$, $\Delta \theta$) &  $\Delta s_1^{\mu}$ \\
\hline Bound &  $K^{\phi r}$,  $K^{\theta r}$ &  $K^{\phi_S r}$ \\
\hline
\end{tabular}
\caption{The table shows a parallel between observables for the scattering and bound case for generic spin orientations. }
\label{tab:boundvsunbound}
\end{table}

Now, more than ever, precise theoretical models are needed to understand the inspiral dynamics of compact binaries. Focusing on the the linear in spin dynamics of a spinning probe in Kerr spacetime, we have developed new techniques to compute scattering observables (impulse, spin kick) and bound observables (periastron advance, precession of the orbital plane and a newly defined internal spin precession) using the worldline formalism.

We first described how to obtain the radial (and polar) momentum for the spinning probe by choosing an appropriate tetrad adapted to the geodesic motion, recovering a recent result \cite{Witzany:2019nml} with the SUSY $\mathcal{N}=2$ worldline model \cite{Bastianelli:2005vk,Bastianelli:2005uy,Jakobsen:2021zvh}. We then perform, for the first time, an explicit integration of such actions at linear in spin order extending our previous calculation \cite{Gonzo:2023goe}. The new effective action we obtained shows that the radial action can be interpreted as the Hamiltonian for the evolution of our observables: this resonates with the amplitude-action relation recently discovered in the literature \cite{Bern:2021dqo,Kol:2021jjc,Adamo:2022ooq,Damgaard:2021ipf}. 

We then developed a systematic way to extract scattering observables for generic spin orientations in a fully covariant way, generalizing and proving for the first time some conjectural expressions in the amplitude literature. We applied a reduction of Poisson brackets to Dirac brackets using on-shell and transversality constraints, which give ready-to-use expressions for the impulse and spin kick once the radial action (through the exponential S-matrix representation, or the worldline) is extracted. 

We checked the validity of our Dirac brackets by computing, for the first time, the impulse $\Delta v_1^{\mu}$ and the spin kick $\Delta s_1^{\mu}$ in the probe limit up to 6PM and to fourth order in the spin of the Kerr background (i.e., $\mathcal{O}(G_N^6 s_1 s_2^4)$). Using known results from the literature at 2PM and 3PM order in the conservative region, we checked the validity of such brackets also beyond the probe limit. In the future, it would be interesting to use our method to extract scattering observables for generic spin orientations from recent higher spin amplitude calculations \cite{Aoude:2022thd,Aoude:2023vdk,Bern:2022kto,Kosmopoulos:2021zoq}. 

Finally, we discussed bound observables using the action-angle representation inspired by the integrability of the linear in spin Kerr dynamics \cite{rudiger1981conserved,rudiger1983conserved,Gibbons:1993ap,Witzany:2019nml,Ramond:2022vhj,Ramond:2024ozy,Ramond:2024sfp}. We defined the corresponding fundamental frequencies $(K^{\theta r},K^{\phi r},K^{\phi_S r})$, which can be analytically evaluated with our integrated radial and polar action for generic spin orientations. We then provided an explicit PM-expanded expression of the aligned-spin periastron advance in the probe limit up to $\mathcal{O}(G_N^6 s_1 s_2^4)$.

This work opens up new avenues for future investigations. First, we hope that our results will shed new light on how to extract scattering observables for generic spinning binaries in the amplitude or the worldline formalism, in particular through the Dirac bracket procedure discussed here. Regarding bound observables, while the non-perturbative boundary to bound map found in \cite{Gonzo:2023goe} for generic precessing orbits still applies here for the actions, it would be interesting to establish a full correspondence at the level of observables (see table \ref{tab:boundvsunbound}). Furthermore, we are looking forward to a detailed comparison of our fundamental frequencies for generic orbits with the literature, such as the case of spinning particles in Schwarzschild \cite{Witzany:2023bmq}, hoping to import our results directly in the GSF framework. Finally, it is worth investigating whether we can extend our results to 1SF order in an analytic fashion, perhaps building on the recent approaches \cite{Kosmopoulos:2023bwc,Cheung:2023lnj,Khalaf:2023ozy,Adamo:2023cfp}. In particular, this would allow to extend some of our findings to the inclusion of radiative effects.

\begin{acknowledgments}
We thank J. Vines for initial collaboration and for sharing his unpublished notes, which inspired our proof in appendix A and further motivated some of our discussions leading to this work. It is a pleasure to thank F.Alessio, D.Bini, A.Ilderton, D.Kosmopoulos, A.Luna, N.Moynihan, A. Ochirov, D.O'Connell, A. Pound, F. Teng and especially P. Ramond and V. Witzany for various insightful comments and discussions on the manuscript. The work of CS is supported by China Postdoctoral Science Foundation under Grant No. 2022TQ0346, and the National Natural Science Foundation of China under Grant No. 12347146.

\end{acknowledgments}

\appendix

\section{A proof of the radial formula for spinless probe}
\label{sec:proveRadialformula}
In this appendix, we prove the formula for extracting momentum deflection from the scattering radial action in the case of a spinless probe in Kerr background~\footnote{For brevity, we drop the superscripts $G$ and $>$ compared to the definition in the main text.},
\begin{align}
    \label{eq_radialformula}
    \Delta \lambda = \sum_{j=1} \frac{1}{j!} \underbrace{ \{I_r, \{I_r, ..., \{I_r}_{j \text{ times}}, \lambda\} ... \} \},
\end{align}
where $\lambda$ can be any kinematic variable such as $\theta, \phi, v^\mu$ and we omit the subscript of $I_r^{>}$ for brevity.
Our proof takes two steps: first, we derive the relation for the scattering angles $\Delta \theta$, $\Delta \phi$ in the unconstrained system,
\begin{align}
    \label{eq_defDelta}
    \theta_\mathrm{out} = \pi - \theta_\mathrm{in} + \Delta \theta ,
    \quad
    \phi_\mathrm{out} = \pi + \phi_\mathrm{in} + \Delta \phi .
\end{align}
Second, we promote the relation to a covariant form with the Dirac brackets given in \eqref{eq_DiracBrackets}.

Parametrizing the action with the radial coordinate $r$, we can write the Hamilton's principal function as
\begin{gather}
    \label{eq_W}
    S(x^I, P_I; r) = -E t + \phi L
    + \fint_{\theta_\mathrm{in}}^{\theta(r)} p_\theta(\theta) \dd \theta
    + W_r, \\
    W_r(r) = \fint_{\infty}^{r} p_r(r) \dd r,
\end{gather}
where $x^I = (t, \theta, \phi)$, $P_\mu = (E, L, Q)$ are the conserved quantities. Here $p_r(r)$ is obtained by solving the on-shell constraints $g^{\mu\nu} p_\mu p_\nu = - m^2$ and plays the role as the Hamiltonian. A key feature of our proof is that the dependence of the shifted Carter constant $Q$ on $(\theta, p_\theta)$ is separable,
\begin{align}
    \label{eq:QG}
    Q &= p_\theta^2 + f(\theta) \nonumber \\
    &=p_\theta^2+a^2\left(m^2-p_t^2\right) \cos ^2(\theta)+p_\phi^2 \cot ^2(\theta) \,.
\end{align}
Finally, we use the standard Poisson brackets between the angles and the canonical momenta $p_\theta$, $p_\phi$
\begin{align}
    \{ \theta, p_\theta \} = 1,
    \quad
    \{ \phi, p_\phi \} = 1.
\end{align}

\paragraph{Polar angle $\Delta \theta$}
Let us now derive the scattering angle in the polar direction $\Delta \theta$.
The equations of motion for the outgoing polar angle $\theta_\mathrm{out}$ can be derived by taking derivative of \eqref{eq_W} with respect to $Q$
\begin{align}
    \label{eq_thetaeom}
    \hspace{-8pt}\fint_{\theta_\mathrm{in}}^{\theta_{\mathrm{out}}} \frac{\partial  p_\theta}{\partial Q} \dd \theta + W_{r,Q} = 0,\quad
    W_{r,Q}= \frac{\partial W_r(r)}{\partial Q} \bigg|_{r \to \infty}.
\end{align}
We note that the derivative should act on the integrand since $\theta(\tau), r(\tau)$ are taken as independent arguments.
However, for a scattering process, $r(\tau) \to \infty$, we can pull the differential operator out of the radial integral.
To obtain the expression of $\theta_{\mathrm{out}}$ in terms of $W_{r,Q}$, we consider the total derivative of \eqref{eq_thetaeom},
\begin{equation}
    \label{eq_dThetaOut}
    \frac{\partial p_\theta(\theta_{\mathrm{out}})}{\partial Q} \dd \theta_{\mathrm{out}}
    +\dd W_{r,Q} = 0.
\end{equation}
From \eqref{eq:QG}, we have $\frac{\partial p_\theta}{\partial Q} = \frac{1}{2 p_{\theta}}$, so that $\theta_{\mathrm{out}}$ can be formally written as
\begin{equation}
    \Delta \theta = \theta_{\mathrm{out}} - (\pi - \theta_{\mathrm{in}})= -\int_{ W_{r,Q}^{(0)} }^{W^{(0)}_{r,Q} + I_{r,Q}} 2p_\theta\, \dd  W_{r,Q}.
\end{equation}
We define $W_{r}^{(0)} = W_{r} |_{G \to 0}$ as the 0PM order of the radial action, corresponding to the straight-line background.
Therefore, plugging $W_{r,Q}^{(0)} = \partial W_r^{(0)} / \partial Q$ in \eqref{eq_thetaeom} yields $\theta_{\mathrm{out}}= \pi -\theta_{\mathrm{in}}$.
We also define the perturbative radial action $I_r = W_{r} - W_{r}^{(0)}$ and $I_{r,Q} = \partial I_r / \partial Q$. We are interested in a perturbative expansion around $ W_{r,Q}^{(0)} $,
\begin{align}
    \label{eq_Deltatheta}
    \Delta \theta =& \sum_{j=1}^{\infty} \frac{I_{r,Q}^j}{j!}  \bigg(\frac{\partial }{\partial W_{r,Q} } \bigg)^{j-1} (-2p_\theta) \bigg|_{W_{r,Q} = W_{r,Q}^{(0)}} \nonumber \\
    =& \sum_{j=1}^{\infty} \frac{I_{r,Q}^j}{j!}  \bigg(-2 p_\theta \frac{\partial }{\partial \theta} \bigg)^{j-1} (-2p_\theta) \bigg|_{\theta = \pi - \theta_{\mathrm{in}}}.
%    =& \sum_{j=1}^{\infty} \frac{I_{r,Q}^j}{j!} \bigg(2 p_\theta \frac{\partial }{\partial \theta} \Big|_{Q} \bigg)^{j-1} (2p_\theta) \bigg|_{\theta =  \theta_{\mathrm{in}}},
\end{align}
We stress that the derivative is taken with $Q$ being invariant, which will be crucial for deriving a Poisson bracket representation.
To rewrite the above equation in terms of Poisson brackets, we note that the dependence of the radial action $I_r$ on $(\theta, p_{\theta})$ is solely via $Q$.
Consequently, the nested Poisson brackets in \eqref{eq_radialformula} can be simplified as
\begin{gather}
    \label{eq_Ir2Q}
%    \{ \dots \{\theta, \underbrace{I_r\}, I_r\} \dots I_r\}}_{j \text{ times}} = I_{r,Q}^j \{ \dots \{\theta, \underbrace{Q\}, Q\} \dots Q\}}_{j \text{ times}}
    \underbrace{ \{I_r, ..., \{I_r}_{j \text{ times}}, \theta\} ... \} \} = I_{r,Q}^j \underbrace{ \{Q, ..., \{Q}_{j \text{ times}}, \theta\} ... \} \}
\end{gather}
For an arbitrary function $F(\theta, p_\theta)$, we have explicitly 
\begin{align}
    \{Q, F\} = \frac{\partial Q}{\partial \theta}\Big|_{p_{\theta}} \frac{\partial F}{\partial p_{\theta}} - \frac{\partial Q}{\partial p_{\theta}} \frac{\partial F}{\partial \theta}\Big|_{p_{\theta}} ,
\end{align}
Here the derivative with respect to $\theta$ is taken by keeping $p_\theta$ invariant.
A key observation is that, with the change of variables $(\theta, Q) \leftrightarrow (\theta, p_\theta)$, we have
\begin{align}
    \label{eq_DtoPoisson}
    \frac{\partial }{\partial \theta} \Big|_{Q} = \frac{\partial }{\partial \theta} \Big|_{p_\theta} + \frac{\partial p_\theta}{\partial \theta} \Big|_{Q} \frac{\partial }{\partial p_\theta}
\end{align}
As it is mentioned previously, that the dependence of $Q$ on $p_\theta, \theta$ is separable plays a crucial role here,
\begin{align}
    -2 p_\theta \frac{\partial p_\theta}{\partial \theta} \bigg|_Q \!
    = -\frac{\partial p_\theta^2}{\partial \theta}\bigg|_Q \!
    = \frac{\partial f(\theta) }{\partial \theta} \bigg|_Q \!
    = \frac{\partial f(\theta) }{\partial \theta} \bigg|_{p_\theta} \!\!
    =  \frac{\partial Q}{\partial \theta}\bigg|_{p_\theta},
\end{align}
which allows us to rewrite 
\begin{align}
    \label{eq_PBtoD}
    \{Q, F\} = -2 p_\theta \frac{\partial F}{\partial \theta} \bigg|_{Q}.
\end{align}
This allows to rewrite \eqref{eq_Deltatheta} as 
\begin{align}
    \label{eq_thetaofchi}
%    \Delta \theta =& \sum_{j=1}^{\infty} \frac{I_{r,Q}^j }{j!} \{ \dots \{2p_\theta(\theta_{\mathrm{in}}), \underbrace{ Q\}, \dots Q \}}_{(j-1) \text{ times}} \nonumber \\
%    =&\sum_{j=1}^{\infty} \frac{I_{r,Q}^j }{j!} \{ \dots \{\theta_{\mathrm{in}}, \underbrace{ Q\},Q\}, \dots Q \}}_{j \text{ times}} 
    \Delta \theta =& \sum_{j=1}^{\infty} \frac{I_{r,Q}^j }{j!} \underbrace{ \{Q, ..., \{Q}_{(j-1) \text{ times}}, -2 p_\theta\} ...  \} \big|_{\theta = \pi -\theta_{\mathrm{in}}} \nonumber \\
    =&\sum_{j=1}^{\infty} \frac{I_{r,Q}^j }{j!} \underbrace{ \{Q, ..., \{Q, \{Q}_{j \text{ times}}, \theta\} ... \} \} \big|_{\theta = \pi -\theta_{\mathrm{in}}} 
\end{align}
Finally, with \eqref{eq_Ir2Q}, we claim to prove \eqref{eq_radialformula} for $\lambda = \theta$.

\paragraph{Azimuthal angle $\phi$}
The azimuthal deflection angle is obtained from taking derivative of the characteristic function \eqref{eq_W} with respect to $L$,
\begin{align}
    \label{eq_phieom}
    \phi_{\mathrm{out}} - \phi_{\mathrm{in}} = - \frac{\partial W_r}{\partial L} \Big|_{Q} - \fint_{\theta_\mathrm{in}}^{\theta_{\mathrm{out}}} \frac{\partial p_\theta}{\partial L} \Big|_{Q} \dd \theta
    + \text{const}
\end{align}
Importantly, the polar integrand can be simplified by exploiting the definition of the Carter constant \eqref{eq:QG}
\begin{align}
    2 p_\theta\frac{\partial p_\theta}{\partial L}\Big|_{Q}
    = \frac{\partial p_\theta^2}{\partial L}\Big|_{Q}
    = - \frac{\partial f(\theta)}{\partial L}\Big|_{Q}
    = - \frac{\partial Q}{\partial p_\phi}\Big|_{p_\theta},
\end{align}
where in the last equal sign we used $p_\phi = L$.
Furthermore, with the differential equation between $\theta_{\mathrm{out}}$ and the radial action~\eqref{eq_dThetaOut}, we have
\begin{align}
    \label{eq_phiWr}
    \phi_{\mathrm{out}} - \phi_{\mathrm{in}} = - \frac{\partial W_r}{\partial L} \Big|_{Q}
    - \fint_{0}^{W_{r,Q} } \frac{\partial Q}{\partial p_\phi} \Big|_{p_\theta} \dd W_{r,Q}
    + \text{const}
\end{align}
We are interested in the perturbation around a straight-line background, where $\phi_{\mathrm{out}} - \phi_{\mathrm{in}} = \pi$ if we plug $W_{r,Q}= W_{r,Q}^{(0)}$ in \eqref{eq_phiWr}.
Subtracting this unperturbed background from \eqref{eq_phiWr}, we obtain
\begin{align}
    \label{eq_phiWr2}
    \Delta \phi = - \frac{\partial I_r}{\partial L} \Big|_{Q}
    - \int_{W_{r,Q}^{(0)}}^{W_{r,Q}^{(0)} + I_{r,Q}} \frac{\partial Q}{\partial p_\phi} \Big|_{p_\theta} \dd W_{r,Q}.
\end{align}
The perturbative expansion of the last term is then
\begin{align}
%    \Delta \phi =& - \frac{\partial I_r}{\partial L}\Big|_{Q}
     & - \sum_{j=1}^{\infty} \bigg[ \frac{I_{r,Q}^j}{j!} \Big(\frac{\partial }{\partial W_{r,Q} } \Big|_{Q} \Big)^{j-1} \Big( \frac{\partial Q}{\partial p_\phi}\Big|_{p_\theta} \Big) \bigg] \bigg|_{W_{r,Q} = W_{r,Q}^{(0)}} \nonumber \\
    =& %- \frac{\partial I_r}{\partial L}\Big|_{Q}
      - \sum_{j=1}^{\infty} \bigg[ \frac{I_{r,Q}^j}{j!} \Big({-}2 p_\theta \frac{\partial }{\partial \theta}\Big|_{Q} \Big)^{j-1} \Big( \frac{\partial Q}{\partial p_\phi}\Big|_{p_\theta} \Big) \bigg] \bigg|_{\theta = \pi - \theta_{\mathrm{in}}} ,
%    =& 
%    - \sum_{j=1}^{\infty} \bigg[ \frac{I_{r,Q}^j}{j!} \Big(2 p_\theta \frac{\partial }{\partial \theta}\Big|_{Q} \Big)^{j-1} \Big( \frac{\partial Q}{\partial p_\phi}\Big|_{p_\theta} \Big) \bigg] \bigg|_{\theta = \theta_{\mathrm{in}}} ,
\end{align}
where we used the relation \eqref{eq_dThetaOut}.
Rewriting the derivative in terms of Poisson bracket as \eqref{eq_PBtoD}, we find 
\begin{align}
%    \Delta \phi = - \frac{\partial I_r}{\partial L} \Big|_{Q} 
%    - \sum_{j=1}^{\infty}  \frac{I_{r,Q}^j}{j!} \Big\{ \dots \Big\{\frac{\partial Q}{\partial p_\phi}\Big|_{p_\theta}, \underbrace{ Q\Big\}, Q\Big\}, \dots Q \Big\}}_{(j-1) \text{ times}} ,
    \Delta \phi = - \frac{\partial I_r}{\partial L} \Big|_{Q} 
    - \sum_{j=1}^{\infty}  \frac{I_{r,Q}^j}{j!}    
     \underbrace{ \Big\{Q, ..., \Big\{Q}_{(j-1) \text{ times}}, \frac{\partial Q}{\partial p_\phi}\Big|_{p_\theta} \} ... \} \Big|_{\theta = \pi -\theta_{\mathrm{in}}}.
\end{align}
Furthermore, the Poisson brackets for the $\phi$ angle and the radial action is
\begin{align}
%    \{\phi_\mathrm{in}, I_r \}
%    = \frac{\partial I_r}{\partial p_{\phi}} \Big|_{p_\theta}
%    = \frac{\partial I_r}{\partial L} \Big|_{Q} + I_{r,Q} \frac{\partial Q}{\partial p_{\phi}}
    \{I_r, \phi\}
    = -\frac{\partial I_r}{\partial p_{\phi}} \Big|_{p_\theta}
    = -\frac{\partial I_r}{\partial L} \Big|_{Q} - I_{r,Q} \frac{\partial Q}{\partial p_{\phi}}
\end{align}
Thanks to the axial symmetry, the radial action doesn't depend on $\phi$ explicitly, so that the higher-order nested Poisson brackets has contribution only from the $(\theta, p_\theta)$ components.
Noting that
\begin{align}
    \bigg\{ I_r, \frac{\partial I_r}{\partial L}\bigg|_{Q} \bigg\} = 0, \quad
    \{ I_r, I_{r,Q} \} =0,
\end{align}
and utilizing\eqref{eq_PBtoD} we can prove \eqref{eq_radialformula} for the azimuthal deflection angle $\lambda = \phi$,
\begin{align}
    \Delta \phi = {- \frac{\partial I_r}{\partial L}\Big|_Q} + \sum_{j=1}^{\infty}  \frac{I_{r,Q}^j}{j!}    
    \underbrace{ \{Q, ..., \{Q}_{j \text{ times}}, \phi \} ... \} \big|_{\theta = \pi -\theta_{\mathrm{in}}}.
\end{align}
In the equatorial limit, we have $Q=0$ and the second term in \eqref{eq_phiWr} vanishes, recovering the familiar result.

\paragraph{Covariant form}
A direct consequence of the Poisson bracket representation of the deflection angles \eqref{eq_radialformula} is that an arbitrary function $g(\theta, \phi)$ satisfy
\begin{align}
    &g(\theta + \Delta \theta, \phi + \Delta \phi) - g(\theta, \phi) = \\ 
    &\sum_{j=1} \frac{1}{j!}\underbrace{\{I_r, ..., \{I_r, \{I_r}_{j \text{ times}}, g(\theta, \phi)\}...\}\} \big|_{\theta = \pi -\theta_{\mathrm{in}}, \phi = \pi + \phi_{\mathrm{in}}}. \nonumber
\end{align}
Taking $g$ as the velocity $v^\mu(\theta, \phi)$, we immediately obtain
\begin{align}
    \label{eq_DeltapinPB}
%    \Delta p^\mu = \sum_{j=1} \frac{1}{j!}\{ \dots \{p^\mu(\theta_{\mathrm{in}}, \phi_{\mathrm{in}}), \underbrace{I_r\}, I_r\} \dots I_r\}}_{j \text{ times}}.
    \Delta v^\mu = \sum_{j=1} \frac{1}{j!} \underbrace{\{I_r, ..., \{I_r}_{j \text{ times}}, v_{\mathrm{in}}^\mu \}...\} ,
\end{align}
where we have used $v_{\mathrm{in}}^\mu = v^\mu(\pi - \theta_{\mathrm{in}}, \pi + \phi_{\mathrm{in}})$.
Furthermore, one can easily check that the Dirac brackets in the main text \eqref{eq_DiracBrackets} fully reproduce all Poisson brackets of $\{\theta, p_\theta, \phi, p_\phi\}$, including the vanishing ones. Therefore, we can simply promote the unconstrained Poisson brackets in \eqref{eq_DeltapinPB} to the covariant Dirac brackets. This completes our proof of \eqref{eq_nestedDB2} for a spinless probe in Kerr spacetime.

\section{The amplitude-action relation in the spherical and spheroidal basis}
\label{app:ampl_action}

In this appendix, we explain how different choices in the partial wave decomposition of a 4-point conservative amplitude for spinning particles {lead} to different {representations of the amplitude-action relation}. We focus on the probe limit of a spinless particle in Kerr, for simplicity, but this suggests a general principle valid for arbitrary interactions of spinning particles representing Kerr black hole binaries.

\begingroup\color{black}
In the nonrotating limit $a=0$, the conventional partial wave decomposition in spherical harmonics $Y_{\ell m}$ reads:
\endgroup
\begin{align}
\label{eq:partialwave_spherical}
\mathcal{M}_4 &= \sum_{\ell=0}^{+\infty} (2 \ell + 1) P_\ell(\cos(\Theta)) \frac{e^{2 i \delta^{Y}_\ell} - 1}{{2}} \\
&= 2 \pi \sum_{\ell=0}^{+\infty} \sum_{m = -\ell}^{+\ell} Y_{\ell m}(\theta) Y_{\ell m}(\gamma)  e^{i m {(\phi - \pi/2)}} \begingroup\color{black}
(e^{2 i \delta^{Y}_{\ell}} - 1)
\endgroup{,}  \nonumber
\end{align}
where $\delta^{Y}_\ell$ is the phase shift{,} and in the last line we have taken $\Theta$ to be the angle on the sphere between the scattering direction $(\theta, \phi)$ and the antipode at $(\gamma,\pi/2)$  \cite{Glampedakis:2001cx}{:}
\begin{align}
\cos(\Theta) = \cos(\theta) \cos(\gamma) + \sin(\theta) \sin(\gamma) \sin(\phi) \,.
\end{align}

There is an alternative partial wave representation of the 4-point amplitude in spheroidal harmonics $\mathcal{S}_{\ell m}^{a}$ adapted to the Kerr spin axis $\vec{a} = (0,0,a)$ \cite{Glampedakis:2001cx,Bautista:2021wfy}{:}
\begin{align}
\mathcal{M}_4 &= 2 \pi \sum_{\ell=0}^{+\infty} \sum_{m = -\ell}^{+\ell} \mathcal{S}_{\ell m}^{a}(\theta) \mathcal{S}^{a}_{\ell m}(\gamma) e^{i m {(\phi - \pi/2)}} (e^{2 i \delta^{\mathcal{S}}_{\ell m}} - 1)\,,
\label{eq:partialwave_spheroidal}
\end{align}
which reduces to \eqref{eq:partialwave_spherical} in the spinless limit $\mathcal{S}^{a}_{\ell m} \stackrel{a \to 0}{\to} Y_{\ell m}${,} but it is generally a more useful basis to study the scattering problem in Kerr because it gives separable equations.

The derivation of the radial action representation for the spinless{, spherically symmetric} case was {rigorously} shown in \cite{FORD1959259,Kol:2021jjc}{;} see also \cite{Bern:2021dqo,Damgaard:2021ipf,Adamo:2022ooq}. The spinning case is more subtle, given for a spin $s$-particle there are naively $(2 s + 1)$ states for the quantum description{,} and therefore we should talk about a matrix in spin space. Nevertheless, using spin coherent states as in \cite{Bern:2020buy,Aoude:2021oqj}{,} we do expect a localization in terms of classical spin vectors $s_1^{\mu}$, $s_2^{\mu}${,} which should combine into a single scalar function, at least for Kerr black holes. Coming back to our problem of a spinless particle in Kerr, these findings suggest that by taking the classical limit of  \eqref{eq:partialwave_spheroidal} we land on {the following amplitude-action relation:}
\begingroup\color{black}
\begin{align}
\sum_{m=-\ell}^{+\ell}\mathcal{S}_{\ell m}^{a}(\theta)\mathcal{S}_{\ell m}^{a}(\gamma)
 e^{im(\phi-\pi/2)}e^{2i\delta^{\mathcal S}_{\ell m}}
&\stackrel{\hbar\to0}{\sim}
 e^{\frac{i}{\hbar}(I_r^{>}+I_\theta^{>}+L\pi)},
\label{eq:saddle-point}
\end{align}

\endgroup%
where the non-trivial polar action $I^{>}_{\theta}$ in the spheroidal basis decomposition {is expected to arise} from the WKB approximation of $\mathcal{S}^{a}_{\ell m}${; see} \cite{Hod:2012dtv} {for a WKB analysis of spheroidal harmonics.} It would be interesting to put \eqref{eq:saddle-point} on a more rigorous footing{, including the matching of angular and free reference phases}, but this is beyond the scope of this work.

Interestingly, scattering observables like deflection angles can be obtained in different ways from the actions. Indeed, in \cite{Gonzo:2023goe} we computed the azimuthal angle $\Delta\phi$ from
\begin{align}
\Delta\phi + \pi = -\frac{\partial (I^{>}_r + I^{>}_{\theta})}{\partial L}{,} 
\label{eq:angle_basis1}
\end{align}
\begingroup\color{black}
where $L$ is conjugate to $\phi$. In this work we instead use the equivalent canonical bracket representation:
\endgroup
\begin{align}
\Delta\phi + \pi = \{I^{>}_r, \phi\} + \frac{1}{2 !} \{I^{>}_r,\{I^{>}_r, \phi \}\}  + \ldots \,,
\label{eq:angle_basis2}
\end{align}
which gives an equivalent answer. 

\section{Post-Minkowskian expansion of the scattering radial action}

We report here the explicit results for the Post-Minkowskian expansion of the (scattering)  radial action $I_r^{>}$ in terms of the unit-mass variables $\bar{E},\mathcal{E},l,l_Q,a, s_{\parallel}$, see table \ref{tab:IrG} for $I_r^{\mathrm{G}>}$ and \ref{tab:IrS} for $I_r^{\mathrm{S}>}$. 
These results are also provided in the ancillary file. 
There are many interesting patterns we observe from the expressions for the radial action (except for $\mathcal{O}(G^1 a^0)$):
\begin{enumerate}
    \item Most significantly, the dependence of the energy $\bar{E}$ and the angular momentum and Carter's constant $l, l_Q$ factor out;
    \item For the energy dependence, after factoring out proper powers of $\calE$, the radial action depends on some polynomials in $\bar{E}$ as commonly observed in binary dynamics;
    \item Less trivially, after factoring some powers of $l, l_Q$, the radial action depends on some simple polynomials in $l^2/l_Q^2$.
\end{enumerate}
These patterns implies a hypergeometric representation of the radial action that manifest PM expansion and the factorization between $E$ and $l,l_Q$ dependence, simpler than the one we found in terms of Lauricella's $F_D$.

\begin{table*}[!h]
\begin{center}
    \caption{$I_r^{\mathrm{G} >} /m$ up to $\calO(G^6 s_{\parallel}^0 a^4)$.}\label{tab:IrG}
\resizebox{2\columnwidth}{!}{\begin{tabular}{|c|c|c|c|}
    \hline
    $I_r^{\mathrm{G} >} /m$ &
    $a^0$ &
    $a^1$ &
    $a^2$  \\
    \hline
    $G^1M^1$ &
    $\frac{2\left(1-2 \bar{E}^2\right) \log \left({l_Q}/{(2{\sqrt{\calE}})}\right) -2 {\calE}}{{\sqrt{\calE}}}$ &
    $-\frac{4 {\sqrt{\calE}} \bar{E} l}{l_{Q}^2}$ &
    $-\frac{{\sqrt{\calE}} \left(2 \bar{E}^2-1\right) \left(l_Q^2-2 l^2\right)}{l_Q^4}$ \\
    \hline
    $G^2M^2 \pi$ &
    $\frac{3 \left(5 \bar{E}^2-1\right)}{4 l_Q}$ &
    $\frac{\bar{E} \left(3-5 \bar{E}^2\right) l}{l_Q^3}$ &
    $-\frac{\left(95 \bar{E}^4-102 \bar{E}^2+15\right) \left(l_Q^2-3 l^2\right) }{32 l_Q^5}$ \\
    \hline
    $G^3M^3$ &
    $\frac{64 \bar{E}^6-120 \bar{E}^4+60 \bar{E}^2-5}{3 {\calE^{3/2}} l_Q^2}$ &
    $-\frac{4 \bar{E} \left(16 \bar{E}^4-20 \bar{E}^2+5\right) l}{{\sqrt{\calE}} l_Q^4}$ &
    $-\frac{ \left(128 \bar{E}^6-216 \bar{E}^4+96 \bar{E}^2-7\right) \left(l_Q^2-4 l^2\right)}{3 {\sqrt{\calE}} l_Q^6}$ \\
    \hline
    $G^4M^4 \pi$ &
    $\frac{35 \left(33 \bar{E}^4-18 \bar{E}^2+1\right)}{64 l_Q^3}$ &
    $-\frac{21 \bar{E} \left(33 \bar{E}^4-30 \bar{E}^2+5\right) l}{8 l_Q^5}$ &
    $-\frac{21 \left(759 \bar{E}^6-1005 \bar{E}^4+325 \bar{E}^2-15\right) \left(l_Q^2-5 l^2\right)}{256 l_Q^7}$ \\
    \hline
    $G^5M^5$ &
    $\frac{1792 \bar{E}^{10} {-}5760 \bar{E}^8{+}6720 \bar{E}^6{-}3360 \bar{E}^4+630 \bar{E}^2-21}{10 {\calE^{5/2}} l_Q^4}$ &
    $-\frac{4 \bar{E} \left(896 \bar{E}^8-2304 \bar{E}^6+2016 \bar{E}^4-672 \bar{E}^2+63\right) l}{3 {\calE^{3/2}} l_Q^6} $ &
    $-\frac{ \left(13568 \bar{E}^{10}-40320 \bar{E}^8+43200 \bar{E}^6-19680 \bar{E}^4+3330 \bar{E}^2-99\right) \left( l_Q^2-6 l^2\right)}{15 {\calE^{3/2}} l_Q^8}$ \\
    \hline
    $G^6M^6 \pi$ &
    $\frac{231  \left(221 \bar{E}^6-195 \bar{E}^4+39 \bar{E}^2-1\right)}{256 l_Q^5}$ &
    $-\frac{495 \bar{E} \left(221 \bar{E}^6-273 \bar{E}^4+91 \bar{E}^2-7\right) l}{64 l_Q^7} $ &
    $-\frac{165 \left(33371 \bar{E}^8-54236 \bar{E}^6+26754 \bar{E}^4-4060 \bar{E}^2+91\right) \left({l_Q^2}-7 l^2\right)}{4096 l_Q^9}$ \\
    \hline
\end{tabular}}
\\
\resizebox{2\columnwidth}{!}{\begin{tabular}{|c|c|c|}
    \hline
    $I_r^{\mathrm{G} >} /m$ &
    $a^3$ &
    $a^4$ \\
    \hline
    $G^1M^1$ &
    $-\frac{4 {\calE^{3/2}} \bar{E} l \left(4 l^2-3 l_Q^2\right) }{3 l_Q^6}$ &
    $\frac{{\calE^{3/2}} \left(2\bar{E}^2 -1\right) \left(8 l^4-8 l_Q^2 l^2+l_Q^4\right)}{2 l_Q^8}$ \\
    \hline
    $G^2M^2 \pi$ &
    $-\frac{3 {\calE} \bar{E} \left(9 \bar{E}^2-5\right) l \left(5 l^2-3 l_Q^2\right) }{8 l_Q^7}$ &
    $\frac{{\calE} \left(239 \bar{E}^4-250 \bar{E}^2+35\right) \left(35 l^4-30 l_Q^2 l^2+3 l_Q^4\right)}{256 l_Q^9}$ \\
    \hline
    $G^3M^3$ &
    $-\frac{8 {\sqrt{\calE}} \bar{E} \left(80 \bar{E}^4-92 \bar{E}^2+21\right) l \left(2 l^2-l_Q^2\right)}{3 l_Q^8}$ &
    $\frac{{\sqrt{\calE}} \left(64 \bar{E}^6-104 \bar{E}^4+44 \bar{E}^2-3\right) \left(16 l^4-12 l_Q^2 l^2+l_Q^4\right) }{l_Q^{10}}$ \\
    \hline
    $G^4M^4 \pi$ &
    $-\frac{5 \bar{E} \left(891 \bar{E}^6{-}1575 \bar{E}^4+805 \bar{E}^2-105\right) l \left(7 l^2-3 l_Q^2\right) }{32 l_Q^9}$ &
    $\frac{15 \left(72963 \bar{E}^8 - 162540 \bar{E}^6 + 116690 \bar{E}^4 - 28140 \bar{E}^2 + 1155\right) \left(21 l^4-14 l_Q^2 l^2+l_Q^4\right)}{8192 l_Q^{11}}$ \\
    \hline
    $G^5M^5$ &
    $-\frac{4 \bar{E} \left(640 \bar{E}^8-1536 \bar{E}^6+1248 \bar{E}^4-384 \bar{E}^2+33\right) l  \left(8 l^2-3 l_Q^2\right)}{{\sqrt{\calE}} l_Q^{10}}$&
    $\frac{\left(25856 \bar{E}^{10}-73600 \bar{E}^8+75200 \bar{E}^6-32480 \bar{E}^4+5170 \bar{E}^2-143\right) \left(80 l^4-48 l^2 l_Q^2+ 3 l_Q^4\right)}{30 {\sqrt{\calE}} l_Q^{12}}$\\
    \hline
    $G^6M^6\pi $ &
    $-\frac{385 \bar{E} \left(36907 \bar{E}^8-75348 \bar{E}^6+50778 \bar{E}^4-12516 \bar{E}^2+819\right) l \left(3 l^2- l_Q^2\right)}{1024 l_Q^{11}} $ &
    $\frac{1155 \left(138125 \bar{E}^{10}-342069 \bar{E}^8+297570 \bar{E}^6-106218 \bar{E}^4+13377 \bar{E}^2-273\right) \left(33 l^4-18 l^2 l_Q^2+l_Q^4\right)}{32768 l_Q^{13}} $ \\
    \hline
\end{tabular}}
\end{center}
\end{table*}

\begin{table*}[!h]
\begin{center}
\caption{$I_r^{\mathrm{S} >} /m$ up to $\calO(G^6 s_{\parallel}^1 a^4)$.}\label{tab:IrS}
\resizebox{2\columnwidth}{!}{\begin{tabular}{|c|c|c|c|}
    \hline
        $I_r^{\mathrm{S} >} /m$ &
        $a^0$ &
        $a^1$ &
        $a^2$  \\
    \hline
        $G^1M^1$ &
        $\frac{2 \bar{E}}{{\sqrt{\calE}} l_Q^2}$ &
        $-\frac{4 {\sqrt{\calE}} l }{l_Q^4}$ &
        $-\frac{2 {\sqrt{\calE}} \bar{E} \left(l_Q^2-4 l^2\right)}{l_Q^6}$ \\
    \hline
        $G^2M^2 \pi$ &
        $\frac{3 \bar{E}}{2 l_Q^3}$ &
        $-\frac{9 \left(3 \bar{E}^2 - 1\right) l}{4 l_Q^5}$ &
        $-\frac{3 \bar{E} \left(11 \bar{E}^2-9\right) \left(l_Q^2-5 l^2\right)}{8 l_Q^7}$ \\
    \hline
        $G^3M^3$ &
        $\frac{2 \bar{E} \left(24 \bar{E}^4-40 \bar{E}^2+15\right)}{3 {\calE^{3/2}} l_Q^4} $ &
        $-\frac{40 \left(8 \bar{E}^4-8 \bar{E}^2+1\right) l}{3 {\sqrt{\calE}} l_Q^6}$ &
        $-\frac{4 \bar{E} \left(56 \bar{E}^4-80 \bar{E}^2+25\right) \left(l_Q^2-6 l^2\right) }{3 {\sqrt{\calE}} l_Q^8}$ \\
    \hline
        $G^4M^4 \pi$ &
        $\frac{105 \bar{E} \left(3 \bar{E}^2-1\right)}{16 l_Q^5}$ &
        $-\frac{525 \left(21 \bar{E}^4-14 \bar{E}^2+1\right) l}{64 l_Q^7}$ &
        $-\frac{105 \bar{E} \left(159 \bar{E}^4-170 \bar{E}^2+35\right) \left(l_Q^2-7 l^2\right)}{128 l_Q^9}$ \\
    \hline
        $G^5M^5$&
        $\frac{2 \bar{E} \left(640 \bar{E}^8-1920 \bar{E}^6+2016 \bar{E}^4-840 \bar{E}^2+105\right)}{5 {\calE^{5/2}} l_Q^6}$&
        $-\frac{36 \left(384 \bar{E}^8-896 \bar{E}^6+672 \bar{E}^4-168 \bar{E}^2+7\right) l}{5 {\calE^{3/2}} l_Q^8}$ &
        $-\frac{2 \bar{E} \left(5504 \bar{E}^8-14976 \bar{E}^6+14112 \bar{E}^4-5208 \bar{E}^2+567\right) \left(l_Q^2-8 l^2\right)}{5 {\calE^{3/2}} l_Q^{10}}$ \\
    \hline
        $G^6M^6\pi$ &
        $\frac{1155 \bar{E} \left(39 \bar{E}^4-26 \bar{E}^2+3\right)}{128 l_Q^7}$ &
        $-\frac{8085 \left(143 \bar{E}^6-143 \bar{E}^4+33 \bar{E}^2-1\right) l}{256 l_Q^9}$ &
        $-\frac{1155 \bar{E} \left(1651 \bar{E}^6-2275 \bar{E}^4+861 \bar{E}^2-77\right) \left(l_Q^2-9 l^2\right)}{512 l_Q^{11}}$ \\
    \hline
\end{tabular}}
\\
\resizebox{2\columnwidth}{!}{\begin{tabular}{|c|c|c|}
    \hline
    $I_r^{\mathrm{S} >} /m$ &
    $a^3$ &
    $a^4$ \\
    \hline
    $G^1M^1$ &
    $-\frac{8 {\calE^{3/2}} l \left(2 l^2-l_Q^2\right) }{l_Q^8}$ &
    $\frac{2 {\calE^{3/2}} \bar{E} \left(16 l^4-12 l_Q^2 l^2+l_Q^4\right) }{l_Q^{10}}$ \\
    \hline
    $G^2M^2 \pi$ &
    $-\frac{15 {\calE} \left(17 \bar{E}^2-5\right) l \left(7 l^2-3 l_Q^2\right)}{32 l_Q^9}$ &
    $\frac{45 {\calE} \bar{E} \left(19 \bar{E}^2-15\right) \left(21 l^4-14 l_Q^2 l^2+l_Q^4\right)}{128 l_Q^{11}}$ \\
    \hline
    $G^3M^3$ &
    $-\frac{8 {\sqrt{\calE}} \left(72 \bar{E}^4-64 \bar{E}^2+7\right) l \left(8 l^2-3 l_Q^2\right)}{3 l_Q^{10}}$ &
    $\frac{2 {\sqrt{\calE}} \bar{E} \left(88 \bar{E}^4-120 \bar{E}^2+35\right) \left(80 l^4-48 l_Q^2 l^2+3 l_Q^4\right)}{3 l_Q^{12}}$ \\
    \hline
    $G^4M^4 \pi$ &
    $-\frac{735 \left(441 \bar{E}^6-665 \bar{E}^4+255 \bar{E}^2-15\right) l \left(3 l^2-l_Q^2\right)}{256 l_Q^{11}}$ &
    $\frac{105 \bar{E} \left(4041 \bar{E}^6-7945 \bar{E}^4+4655 \bar{E}^2-735\right) \left(33 l^4-18 l_Q^2 l^2+l_Q^4\right) }{1024 l_Q^{13}}$ \\
    \hline
    $G^5M^5$ &
    $-\frac{48 \left(896 \bar{E}^8-1920 \bar{E}^6+1312 \bar{E}^4-296 \bar{E}^2+11\right) l \left(10 l^2-3 l_Q^2\right)}{5 {\sqrt{\calE}} l_Q^{12}}$ &
    $\frac{12 \bar{E} \left(3712 \bar{E}^8-9600 \bar{E}^6+8544 \bar{E}^4-2952 \bar{E}^2+297\right) \left(40 l^4-20 l^2 l_Q^2+l_Q^4\right)}{5 {\sqrt{\calE}} l_Q^{14}}$ \\
    \hline
    $G^6M^6 \pi$ &
    $-\frac{3465 \left(20163 \bar{E}^8-36036 \bar{E}^6+19866 \bar{E}^4-3444 \bar{E}^2+91\right) l \left(11 l^2-3 l_Q^2\right)}{4096 l_Q^{13}}$ &
    $\frac{1155 \bar{E} \left(87139 \bar{E}^8-193284 \bar{E}^6+143514 \bar{E}^4-39732 \bar{E}^2+3003\right) \left(143 l^4-66 l^2 l_Q^2+3 l_Q^4\right)}{16384 l_Q^{15}}$ \\
    \hline
\end{tabular}}
\end{center}
\end{table*}

\clearpage

\bibliography{references}% Produces the bibliography via BibTeX.

\end{document}